\documentclass{JINST}
\usepackage{epsfig}
\usepackage{subfigure}
\usepackage{here}
\usepackage{graphicx}

\def\figurename{\bf{Fig.\space}}

\title{Advances in ion back-flow reduction in cascaded gaseous electron multipliers incorporating R-MHSP elements.}

\author{A. V. Lyashenko$^a$\thanks{Corresponding
author.}~, A. Breskin$^a$, R. Chechik$^a$, J. F. C. A. Veloso$^b$, J. M. F. Dos Santos$^c$, and F. D. Amaro$^c$\\
\llap{$^a$} Department of Particle Physics, Weizmann Institute of Science,\\
  76100, Rehovot, Israel\\
\llap{$^b$}University of Aveiro,\\
  3810-193 Aveiro, Portugal\\
\llap{$^c$}University of Coimbra,\\
  3004-516 Coimbra,  Portugal\\
  E-mail: \email{alexey.lyashenko@weizmann.ac.il}}

\abstract{A new concept is presented for the reduction of ion
back-flow in GEM-based cascaded gaseous electron multipliers, by
incorporating Micro-Hole \& Strip Plate (MHSP) elements operating in
reversed-bias mode (R-MHSP). About an order of magnitude reduction
in ion back-flow is achieved by diverting back-drifting ions from
their original path. A R-MHSP/2GEM/MHSP cascaded multiplier operated
at total gain of $\sim$1.5*10$^5$ yielded ion back-flow fractions of
1.5*10$^{-3}$ and 4*10$^{-4}$, at drift fields of 0.5 and 0.1 kV/cm,
respectively. A 2R-MHSP/MHSP cascaded multiplier operated at a total
gain of $\sim$10$^5$, yielded an ion back-flow fraction of
3*10$^{-3}$. We discuss the concept for trapping back-flowing ions
in these cascaded multipliers and the relevance to gaseous
photomultiplier and TPC applications; directions for further future
developments are outlined.}


\begin{document}

\section{Introduction.}

During the avalanche process in gaseous detectors, a large number of
ions and photons are produced. In gaseous photomultipliers (GPM),
their impact on the photocathode (PC) has high probability to induce
secondary electron emission (SEE) and to generate secondary
avalanches, known as ion- and photon-feedback. The secondary
avalanches cause significant gain limitations and distortions of the
spatial and temporal information. The photon feedback may be
significantly suppressed by a proper choice of the
electron-multiplier geometrical design and its operation conditions,
and it is practically fully suppressed in cascaded GEM multiplier
configurations \cite{dirk03,buzu00}. The ion backflow is much more
difficult to suppress without affecting the multiplier gain and
detection efficiency, because the ions' and electrons' paths are not
decoupled. The ion backflow to the PC strongly limits the
multiplier's gain; e.g. gains $<$100 were measured in GEM-based
visible-light sensitive GPMs due to the high emission properties of
bialkali photocathodes, while they could attain 10$^6$ values when
blocking the ions with a pulsed ion-gate electrode \cite{breskin05}.
The back-flowing ions have further severe consequences, of physical
and chemical damage to the PC surface and accelerated degradation of
its quantum efficiency (QE) \cite{breskin05,singh00,lyas06}. The ion
backflow is also of great concern in Time Projection Chambers
(TPCs), where the ion-cloud penetration into the drift volume builds
up space-charge and causes rate-dependent dynamic electric-field
distortions, consequently affecting the TPC resolution. Intensive
research has been carried out to reduce this effect by replacing the
standard wire-chamber TPC readout elements by cascaded GEM or
Micromegas multipliers \cite{sauli02}.

Though the demand to reduce ion backflow is common to TPCs and to
GPMs, they differ in their operation conditions; TPCs typically
operate at relatively low gain ($\sim$10$^4$) and with low drift
field above the multiplier ($\sim$0.1-0.2 kV/cm) while GPMs have to
operate at higher gain ($>$10$^5$), to ensure single-photoelectron
sensitivity, and with higher drift fields ($\geq$0.5kV/cm), to
provide efficient photoelectron extraction from the PC
\cite{buzu02}. It should be noted that a 5-fold difference in drift
field values generally implies a corresponding 5-fold higher ion
backflow \cite{bondar03,bachmann99}; the demands for ten-times
higher gain and for full single photoelectron detection efficiency
in GPMs imply further constrains in this case, as will be seen
below.

The fraction of the last-avalanche induced ions flowing back to the
drift volume or to the PC is defined here as \emph{Ion Backflow
Fraction} (IBF). The total number of electrons (and ions) created in
the last avalanche in the cascade, per single initial electron (e.g.
a photoelectron or a single ionization electron), is defined as the
\emph{total gain} of the multiplier. The number of electrons, per
single initial electron, transferred from a given multiplier element
on to a consecutive electrode is defined as the \emph{visible gain}
of that element.

In a single multi-wire or parallel-plate electron multiplier all the
avalanche ions flow back to the cathodes, thus IBF=1. In
cascaded-GEM structures some ions are trapped on GEM surfaces due to
the dipole field within the GEM holes. It was shown \cite{bondar03}
that the IBF in multi-GEM structures does not depend on the gas
filling but rather on the detector's geometry and on the electric
fields configuration. The charge flow of both ions and electrons is
reduced when choosing smaller GEM holes and smaller transfer fields
between successive elements \cite{bondar03}.In cascades comprising
3-4 GEM electrodes, it is possible to reduce the IBF by optimizing
the hole diameter and shape as well as the transfer fields between
the elements. The lowest IBF values reported so far, at total gas
gain $\sim$10$^4$, are of the order of 0.05 and 0.01 at respective
drift fields of 0.5 and 0.1kV/cm \cite{bondar03,bachmann99}.The
lowest ones at higher gains, of $\sim$10$^5$, are $\sim$0.025 and
0.05 for 3 and 4 GEM-multiplier detectors correspondingly, at a
drift field of 0.1kV/cm \cite{buzu02a}. An operation of a 4-GEM
detector with reflective PC on the top most electrode provided an
IBF value of $\sim$0.1 at a gain of 10$^5$-10$^6$ \cite{dirk04a}. It
was recently reported \cite{lotze06,sauli05,roth04a} that a 3-GEM
TPC readout element could be optimized as to have IBF=0.005 at a
drift field of 0.2 kV/cm, by operating it in a highly asymmetric
biasing mode: very high ($\sim$8kV/cm) first and third transfer
fields, very low ($\sim$60V/cm) second transfer field and high bias
voltage ($\sim$350V) on the third GEM. This configuration was
optimized for a total gain of 10$^4$ and it may not be adequate for
higher gains, because some of the fields are already at their
maximum possible value.

The IBF values presently reached in GPMs (with high total gain and
high drift field \cite{bondar03,buzu02a,dirk04a}), of $>$10$^{-2}$,
are not sufficient for their DC operation with visible-light
sensitive PCs; for such operation, considering the SEE probability
of bialkali PCs, IBF$<$10$^{-4}$ is required. Furthermore, TPCs
designed for future particle- and nuclear-physics collider
experiments will operate at high drift fields (0.5-1 kV/cm) and will
also require electron multipliers with IBF$\sim$10$^{-4}$, It is
therefore clear that the current goal is an IBF$\sim$10$^{-4}$ and
better for both GPMs and TPCs modes: in TPCs it will practically
eliminate all ions in the drift region and in single-photon GPMs,
operating at total gains of 10$^5$-10$^6$, it will reduce
photocathode aging and ion-induced secondary avalanches to
acceptable levels \cite{dirk_tes,dirk06}. As long as the IBF is not
reduced to these levels, it constitutes a true obstacle in the
further development of visible-light sensitive GPMs and calls for a
thorough search for novel viable solutions for substantial IBF
reduction.

An effective IBF suppression was already confirmed in TPCs
\cite{nemet83} as well as in multi-GEM based photon detectors
\cite{breskin05,dirk04a}, by means of an ion-gate electrode, which
indeed suppresses the IBF to better than 10$^{-4}$. However, ion
gating often requires an external trigger source and involves
electronic noise problems; it induces dead-time and thus
rate-limitations, restricting its range of applications. We are
therefore searching for ways to reduce the IBF in gaseous electron
multipliers in DC mode.

\begin{figure} [h]
  \begin{center}
  \makeatletter
    \renewcommand{\p@figure}{Fig.\space}
  \makeatother
    \epsfig{file=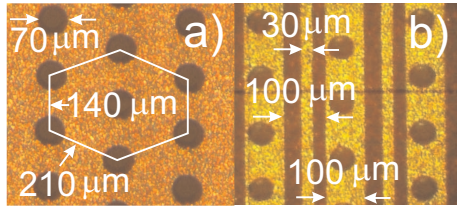, width=6cm}
    \caption{A microscope photograph of an MHSP electrode with 30$\mu$m
anode strips and 100$\mu$m cathode strips. a) top view, b) bottom
view.}
    \label{fig:Fig.1}
  \end{center}
\end{figure}

\begin{figure}[h]%
\begin{center}%
\renewcommand{\thesubfigure}{\thefigure\alph{subfigure}}
\makeatletter
\renewcommand{\p@subfigure}{Fig.\space}
\makeatother %
\subfigure {
    \label{fig:Fig.2:a}
    \includegraphics[width=6.5cm]{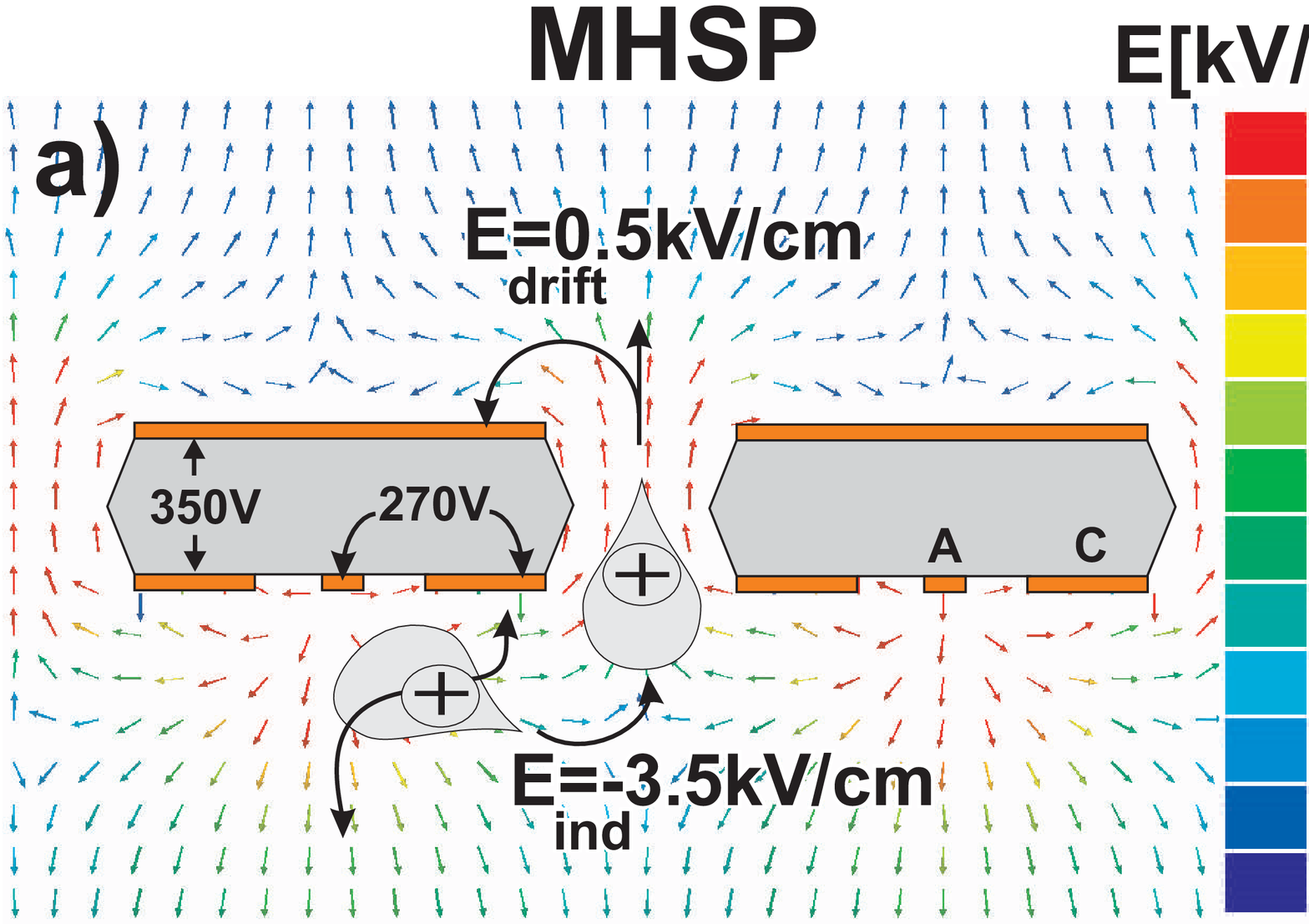}
} \hspace{0.8cm}
\subfigure
{
    \label{fig:Fig.2:b}
    \includegraphics[width=6.3cm]{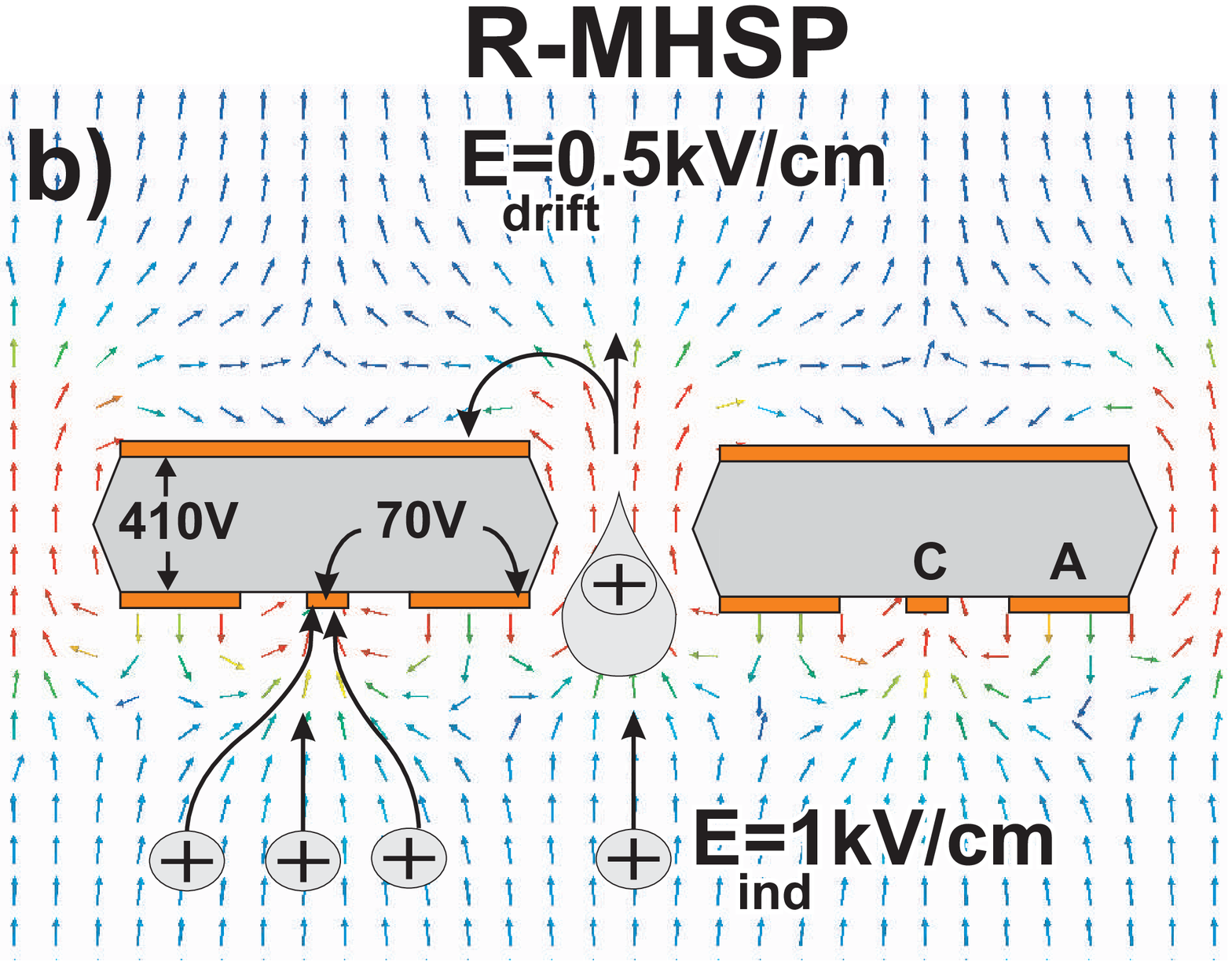}
}
\caption{Schematic views of the operation principles and the
electric-field vectorial maps calculated by MAXWELL software
package, in the vicinity of the electrodes, for: a) normal MHSP b)
reversed-biased R-MHSP. The potentials selected for the field-map
calculations and the color code of the fields are shown in the
figures.}
\label{fig:Fig.2}
\end{center}
\end{figure}

An attempt to do this by modifying the field configuration and
thereby diverting a fraction of the ions away from the PC in a
gaseous photomultiplier was done by replacing the last GEM in the
cascade with a Micro-Hole \& Strip Plate (MHSP) multiplier
\cite{veloso00}.The MHSP, like the GEM, is a hole-multiplier
(\ref{fig:Fig.1}); in addition it has narrow anode strips patterned
on the bottom side, which under positive-voltage biasing provide a
second avalanche multiplication, further to the one occurring within
the electrode's holes. The electric field established between the
anode strips, the cathode strips and the additional cathode plane
(\ref{fig:Fig.2:a}) prevents a large fraction ($~$80\%) of the
final-avalanche ions from back flowing through the hole, as shown in
\cite{maia04}. Moreover, the two-stage multiplication of the MHSP
enables further transfer-field optimization: by setting a small
transfer field above the MHSP, the flow of both ions and electrons
between the GEMs and the MHSP is reduced, but the loss in gain is
recovered by the additional strip multiplication. Using such a
scheme in a gaseous photomultiplier comprising 3-GEMs, a reflective
CsI PC evaporated on the top GEM and an MHSP, the IBF was measured
to be $\sim$0.03 at a total gain of 10$^5$ \cite{maia04,chechik04}.
This clearly demonstrated the potential for IBF reduction when using
the MHSP electrode, by creating different paths for electrons and
ions.

Following this line, it was further proposed \cite{roth04b} to
reverse the roles of the MHSP$'$s anode and cathode strips, in an
attempt to trap ions originating from consecutive multiplication
elements in the cascade, preventing them to flow through the holes.
\ref{fig:Fig.2:a} and \ref{fig:Fig.2:b} show the field configuration
in the immediate proximity of the MHSP hole, for these two different
modes of MHSP operation, defined here as the \emph{normal} (MHSP)
and the \emph{reversed} (R-MHSP) modes (the arrows point at the
field direction, namely to the ion drift direction). Note that in
the normal MHSP mode, the narrow strip-electrodes act as anodes;
they are biased more positive than the broader (cathode) strips
surrounding the holes; in this case double-stage multiplication
takes place within the holes and at the anode strips
(\ref{fig:Fig.2:a}). In the reversed R-MHSP mode, the narrow strips
are biased more negative than the broader ones; charge
multiplication occurs only within the holes while the more-negative
narrow cathode-strips only collect a fraction of the ions
(\ref{fig:Fig.2:b}). The operation mode of the MHSP permits its use
either as a stand-alone detector \cite{veloso00,veloso04} or as the
last element in a cascaded multiplier \cite{maia04}. On the other
hand, the R-MHSP, with its hole-multiplication, can be used anywhere
in the cascade, and especially as a first element, trapping
back-flowing ions from all successive elements (as shown below).

However, one should be aware that in order to have full detection
efficiency of single photoelectrons emitted from the photocathode,
or of ionization electrons radiation-induced within the drift
volume, two conditions have to be fulfilled: 1) the electron's
collection efficiency into the R-MHSP holes, particularly in the
application to single-photon GPMs, has to be close to unity; this
was indeed confirmed for GEMs \cite{bachmann99,richt02} but not yet
for MHSPs, which have slightly more "opaque" hole geometry
(\ref{fig:Fig.1}); 2) the visible gain of the first element in the
cascade should be large enough to ensure full event's detection
efficiency, including in the case of exponential pulse-height
distribution of single photoelectrons; this condition implies a
visible gain of at least 20. The two conditions are of prime
importance, because an electron lost at the first multiplication
element due to inefficient focusing, insufficient multiplication or
inefficient extraction cannot be recovered. Indeed, it was found
that the R-MHSP biasing scheme reduces the extraction efficiency of
the avalanche electrons from the holes towards the next element in
the cascade, thus reducing the visible gain of this multiplier
\cite{veloso05}.

A first attempt to implement the above ideas, cascading two R-MHSPs
with two GEMs \cite{veloso05} resulted in an IBF value of
4*10$^{-3}$, with a drift field of 0.5kV/cm and a total gain of
10$^4$. This study revealed, for the first time, that the ion
trapping also results in a considerable loss of electrons (i.e. loss
of visible gain) - necessitating a careful optimization of the
operation conditions. In addition, this particular experiment was
carried out with a defective R-MHSP electrode, of a limited hole
multiplication; this resulted in a very low visible gain (only 6
electrons in average were transferred into the next GEM
multiplication stage), and did not permit raising the strip voltages
to fully exploit the possibility of ion trapping. Following the
production of better quality MHSP electrodes we continued to study
this avenue of IBF reduction. We will describe below the results
obtained with different cascaded multiplier schemes combining GEM,
MHSP and R-MHSP elements; the optimization of the R-MHSP potentials
permitted reaching higher gains in the first multiplying element,
and resulting in low IBF values with good single-electron detection
efficiencies \cite{breskin05}.

\section{Methodology.}

\begin{figure}[h]%
\begin{center}%
\renewcommand{\thesubfigure}{\thefigure\alph{subfigure}}
\makeatletter
\renewcommand{\@thesubfigure}{\Large\alph{subfigure})}
\renewcommand{\p@subfigure}{Fig.\space}
\renewcommand{\p@figure}{Fig.\space}
\makeatother %
\subfiguretopcaptrue
\subfigure[][] 
{
    \label{fig:Fig.3:a}
    \includegraphics[width=6cm]{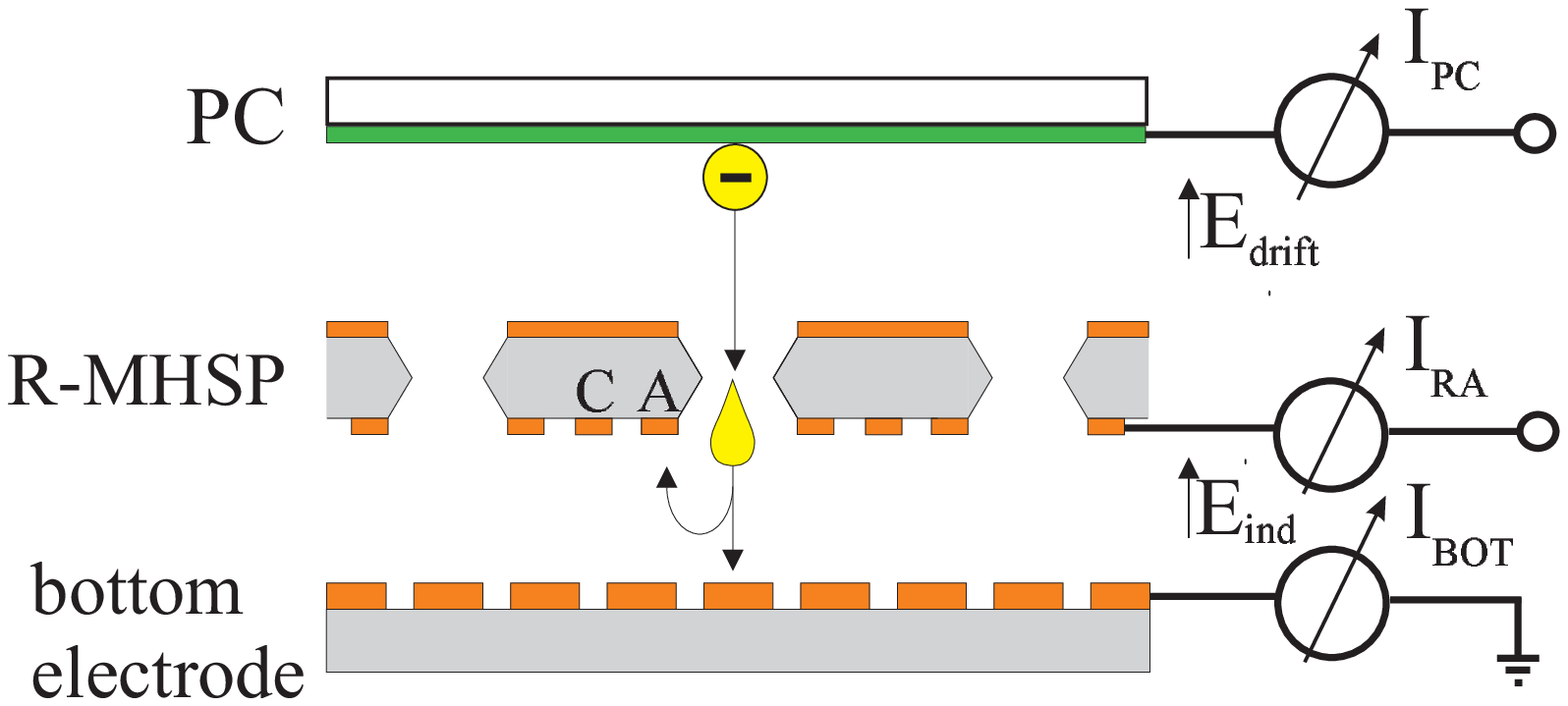}
} \vspace{1cm} \hspace{1cm}
\subfigure[][] 
{
    \label{fig:Fig.3:b}
    \includegraphics[width=6cm]{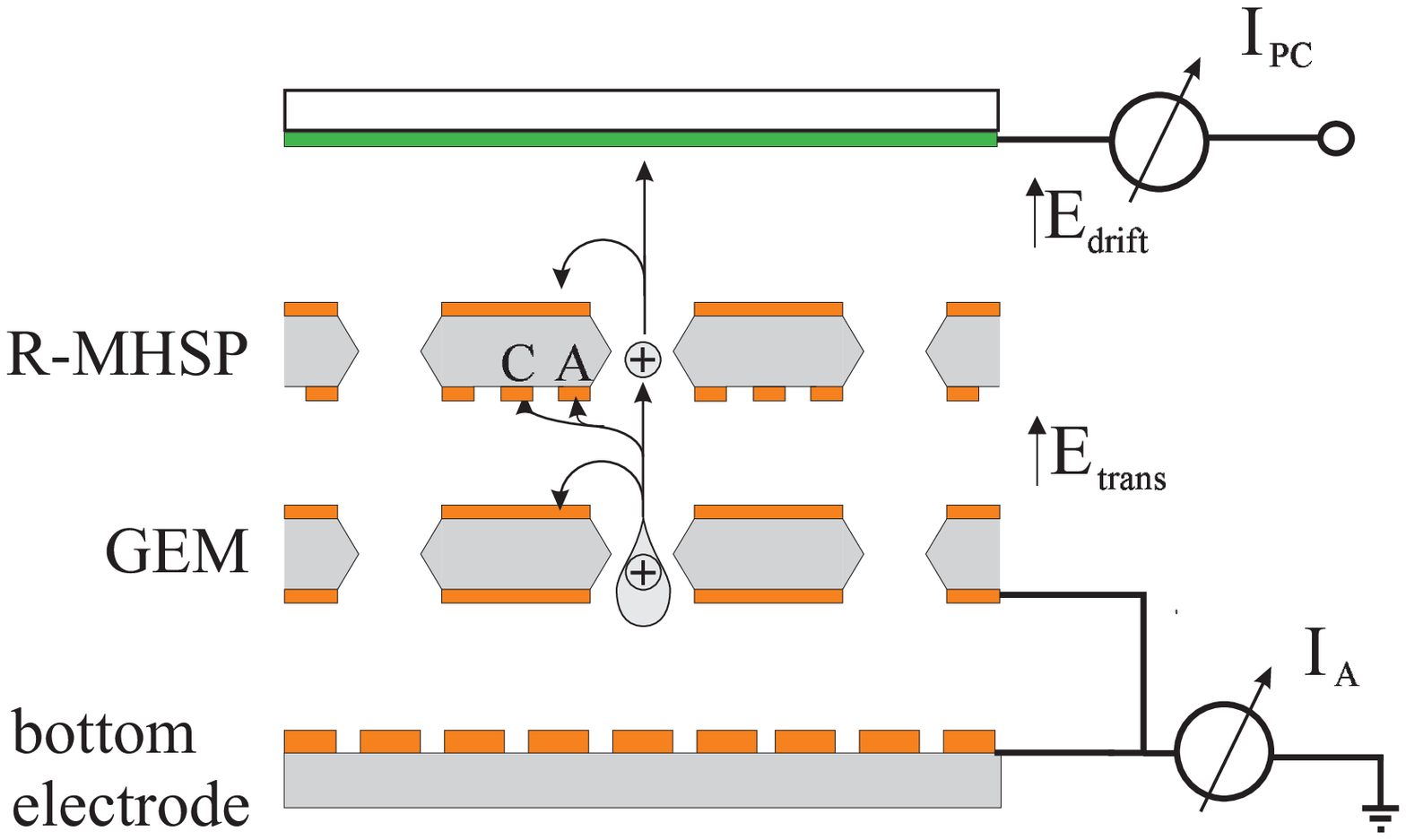}
}
\subfigure[][] 
{
    \label{fig:Fig.3:c}
    \includegraphics[width=6cm]{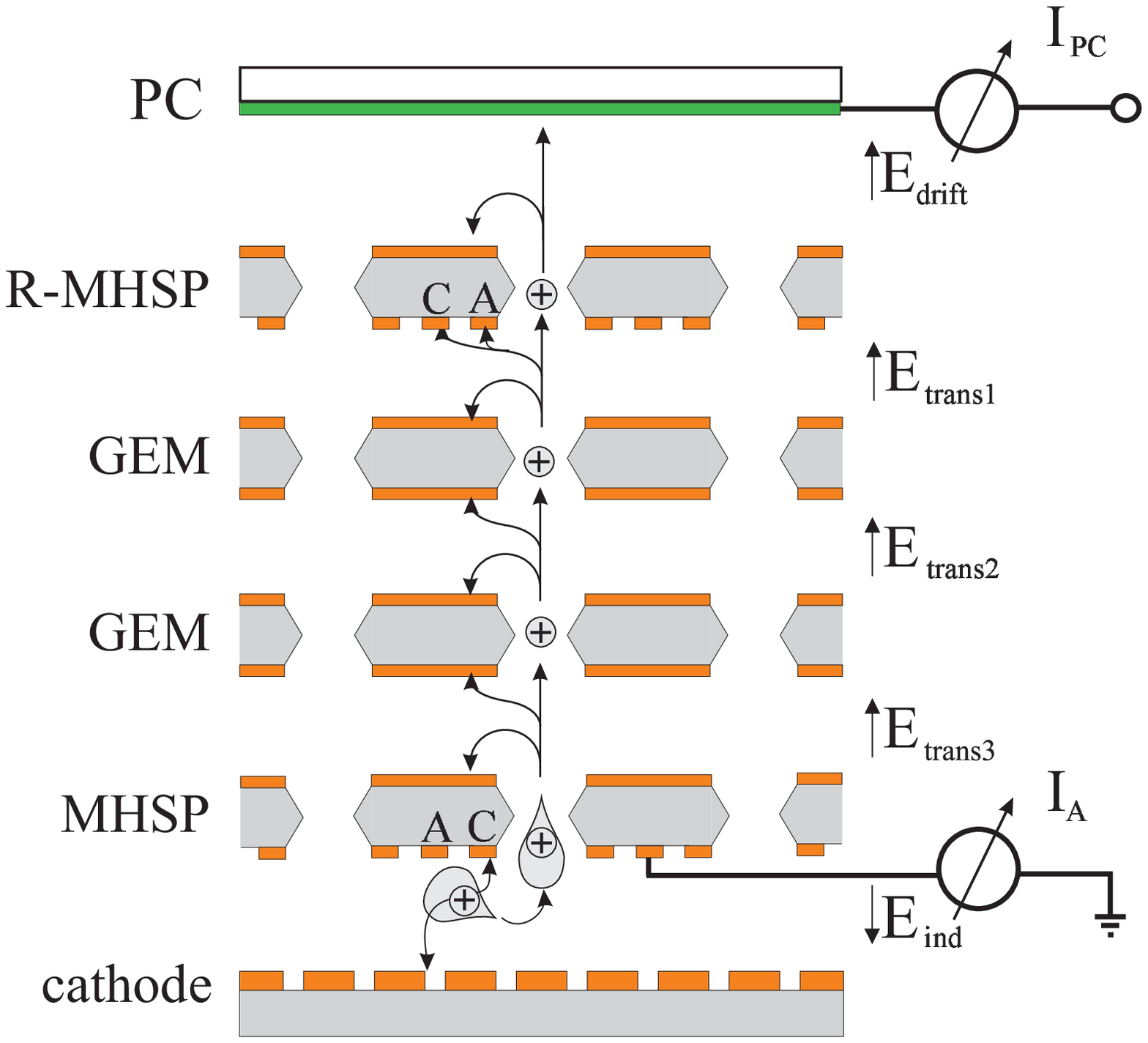}
}\hspace{1cm}
\subfigure[][] 
{
    \label{fig:Fig.3:d}
    \includegraphics[width=6cm]{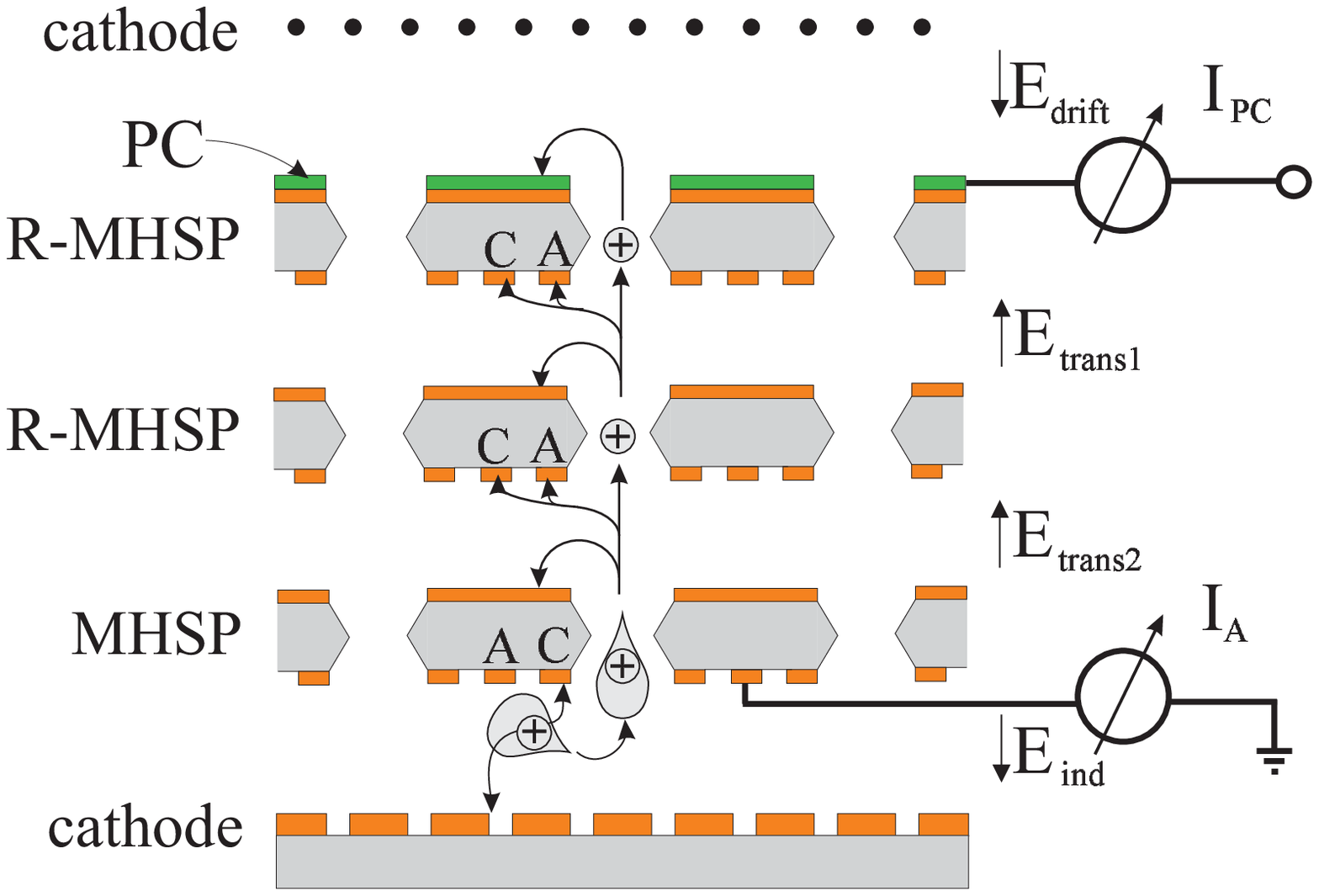}
}
\caption{Four different multiplier configurations and measurement
setups studied in the present work: a) measurement of the total and
visible gains of an R-MHSP; b) measurement of the ion backflow (IBF)
suppression of an R-MHSP; the GEM serves as an ion generator; c) a
cascaded R-MHSP/2GEM/MHSP multiplier coupled to a semi-transparent
photocathode; d) a 2-R-MHSP/MHSP GPM detector with a CsI
photocathode deposited on the top multiplier.}
\label{fig:Fig.3}
\end{center}
\end{figure}

Four multiplier configurations investigated within this work are
shown in \ref{fig:Fig.3}; they combine a UV photocathode as a
single-photoelectron source, a drift space with variable drift
fields, electron multipliers and a readout anode. \ref{fig:Fig.3:a}
depicts a single R-MHSP element coupled to a readout anode; It
permitted investigating and understanding the role of the various
potentials and to optimize their values. \ref{fig:Fig.3:b} shows a
R-MHSP followed by a GEM; the GEM avalanche was used as an "ion
generator" for studying the IBF suppression capability of the
R-MHSP. The cascaded R-MHSP/2GEM/MHSP multiplier is shown in
\ref{fig:Fig.3:c}; this configuration permitted establishing the IBF
value of this cascaded multiplier operating at high total gains, and
using various drift fields corresponding to GPM and TPC operation
conditions. In addition, a detector incorporating 2R-MHSPs followed
by an MHSP is shown in \ref{fig:Fig.3:d}; it represents a
configuration with a reflective radiation converter, e.g. a
UV-photocathode, deposited on the first element in the cascaded GPM.

The MHSP and GEM electrodes, of 28x28 mm$^2$ effective area, were
produced at the CERN printed circuit workshop, from 50$\mu$m thick
Kapton foil with 5$\mu$m copper cladding on both sides. The etched
double-conical 70/50$\mu$m outer/inner diameter GEM holes are
arranged in hexagonal pattern of pitch 140$\mu$m. The MHSP pattern
and dimensions are shown in \ref{fig:Fig.1}. All electrodes were
stretched onto small MACOR frames. The semitransparent PC was 5mm in
diameter; it was made of 300{\AA} thick layer of CsI, evaporated on
a UV transparent window, pre-coated with a 40{\AA} thick Cr film.

The detector elements were mounted within a stainless-steel vessel
evacuated with a turbo-molecular pump to $\sim$10$^7$ Torr prior to
gas filling. The detector was operated with Ar/CH$_{4}$ (95/5) at
760 Torr, under regulated gas flow. The detector was irradiated with
a continuous Ar(Hg) UV-lamp through the window. Each of its
electrodes was biased independently with a CAEN N471A or CAEN N126
power supply.

In all multiplier cascade configurations, the currents on biased
electrodes were recorded as a voltage-drop on a 33M$\Omega$ resistor
with a Fluke 175 voltmeter of 10M$\Omega$ internal impedance. Their
combined resistance was 7.69M$\Omega$ from which the anode current
was calculated. The final avalanche-induced currents following
charge multiplication were always kept below 100 nA by attenuating
the UV-lamp photon flux with absorbers, if necessary, to avoid
charging-up effects. The currents on grounded electrodes were
recorded with a Keithley 485 picoamperemeter.

The IBF, visible gain and total gain were evaluated by recording
currents from the various electrodes in the cascade, as described
below for each of the configurations investigated. In particular: 1)
the total gain in the R-MHSP/2GEM/MHSP detector was calculated from
the ratio of the electron current $I_{A}$ on the anode of the last
multiplying element (MHSP) to the PC photocurrent $I_{PC0}$,
measured in photoelectron collection mode (no gain):
$G_{tot}=I_{A}/I_{PC0}$. The total gain derived from this currents
ratio relates to the average multiplication factor; the number of
electrons in an avalanche (and therefore the number of ions that hit
the PC) is distributed according to a Polya function
\cite{vavra93,vavra96,arley}. 2) The PC current recorded under
multiplication conditions $I_{PC}$, comprises the photocurrent
$I_{PC0}$ and the ion backflow current - the latter is by far
dominant. Therefore, in the R-MHSP/2GEM/MHSP detector the ratio of
$I_{PC}$ to the MHSP anode current recorded under the same
conditions $I_{A}$, provides the IBF$=I_{PC}/I_{A}$. The definitions
are consistent with those of references \cite{maia04,veloso05}.  3)
the visible gain of the first R-MHSP element \ref{fig:Fig.3:a} is
derived from the ratio of anode current $I_{BOT}$, to the PC
photocurrent $I_{PC0}$, measured in photoelectron collection mode:
$G_{VIS}=I_{BOT}/I_{PC0}$.

\section{Results.}

\subsection{Study of a single R-MHSP.}
While the MHSP operation properties are already well established
\cite{maia04}, those of the more recent R-MHSP \cite{veloso05}
required some more basic study. The single R-MHSP study
(\ref{fig:Fig.3:a}, \ref{fig:Fig.3:b}) was designed to yield
understanding of the various potentials$\prime$ role and conditions
for minimal IBF, minimal electron losses and maximal visible gain.
The parameters affecting the R-MHSP$'$s operation are: 1) the hole
bias voltage ($V_{hole}$) which controls the multiplication and the
IBF from the first element; 2) the anode-to-cathode strips voltage
($V_{AC}$), which reduces the visible gain of a single R-MHSP and
reduces IBF from successive elements; 3) the transfer field below
the R-MHSP ($E_{ind}$ in \ref{fig:Fig.3:a} or $E_{trans}$ in
\ref{fig:Fig.3:b}, \ref{fig:Fig.3:c} and \ref{fig:Fig.3:d}), which
affects both the IBF from successive elements and the visible gain
of the R-MHSP.

\begin{figure} [h]
  \begin{center}
  \makeatletter
    \renewcommand{\p@figure}{Fig.\space}
  \makeatother
    \epsfig{file=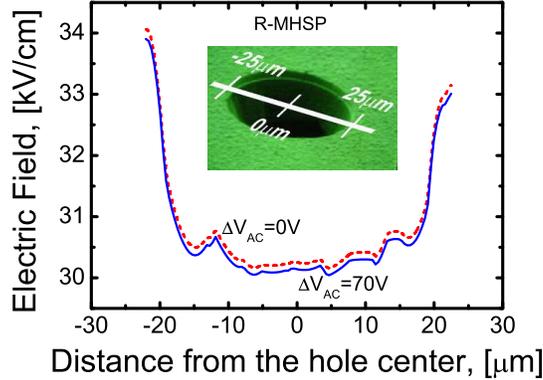, width=8cm}
    \caption{The electric field at the hole-entrance, calculated along
the line shown in the inset. Solid line: ${\Delta}V_{AC}$=70V,
dashed line: ${\Delta}V_{AC}$=0V.}
    \label{fig:Fig.4}
  \end{center}
\end{figure}

For simplicity, and since its influence is similar to that in
cascaded GEMs (i.e. a linear increase of electron extraction
efficiency with this field \cite{dirk04b}), we fixed the field under
the R-MHSP to be 1kV/cm. The drift field $E_{drift}$ was maintained
at 0.5 kV/cm (i.e. typical GPM operation conditions), which with
Ar/CH$_{4}$ (95/5) provides $\sim$70$\%$ efficiency to extract
photoelectrons from the PC \cite{buzu02}.At other $E_{drift}$
values, the IBF increases linearly with the field \cite{bondar03}
and thus a five-fold smaller IBF value is expected with $E_{drift}$=
0.1 kV/cm (i.e. TPC operation conditions).The exact collection
efficiency (namely focusing into the holes) from the semitransparent
PC into the R-MHSP holes, as function of the drift and hole fields
($E_{drift}$, $V_{hole}$), extensively investigated for GEM cascades
\cite{richt02,dirk04b} is a subject of another study, beyond the
scope of the present article. An indication for high collection
efficiency arises from \ref{fig:Fig.2:a}, since there are no field
lines that start at the R-MHSP top face, even though the figure
corresponds to a cut made along the maximal hole distance.
Furthermore, this efficiency can be roughly estimated, by comparison
to studies in GEMs: since the holes of the R-MHSP are arranged in
hexagonal pattern with maximal pitch of 210$\mu$m, it is expected
that the collection efficiency will be not worse than that of GEMs
with identical hole diameter and pitch of 200$\mu$m
\cite{dirk04b,sauli97}, provided the same gas and the same
$V_{hole}$ and $E_{drift}$ are used. The collection efficiency of a
GEM, with a pitch of 200$\mu$m and hole diameter 80$\mu$m and
70$\mu$m, acting as the first cascade element, were studied in
conditions corresponding to operation with semitransparent PC
\cite{sauli97} and reflective PC \cite{dirk04b}. It was found that
in Ar/DME (90/10) with a GEM-MWPC device preceded with a drift
space, with hole voltage of 360V and drift fields in the range of
0.1-1kV/cm, the collection efficiency was of 90-100$\%$
\cite{sauli97}; with a GEM coated with a reflective PC, a collection
efficiency of 90$\%$ was measured with hole voltage above 200V and
drift field 0kV/cm in Ar/CH$_{4}$ (95/5) \cite{dirk04b}. Therefore,
we may expect a collection efficiency ~90$\%$ for our R-MHSP, when
operating with semitransparent PC ($E_{drift}$=0.5kV/cm) at hole
voltage $>$360V and with reflective PC ($E_{drift}$=0kV/cm) at hole
voltage $>$200V.

\begin{figure}[t]%
\begin{center}%
\renewcommand{\thesubfigure}{\thefigure\alph{subfigure}}
\makeatletter
\renewcommand{\p@subfigure}{Fig.\space}
\renewcommand{\p@figure}{Fig.\space}
\makeatother%
\subfiglabelskip=0pt %
\subfigure {
    \label{fig:Fig.5:a}
    \includegraphics[width=7cm]{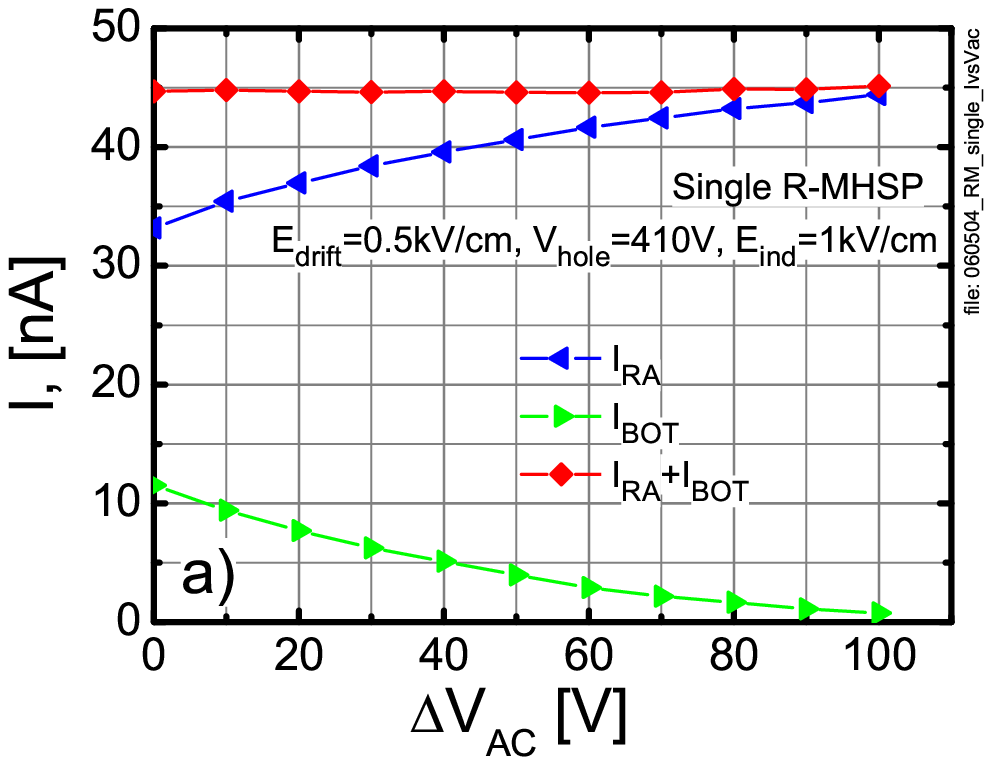}
} \subfigure {
    \label{fig:Fig.5:b}
    \includegraphics[width=7cm]{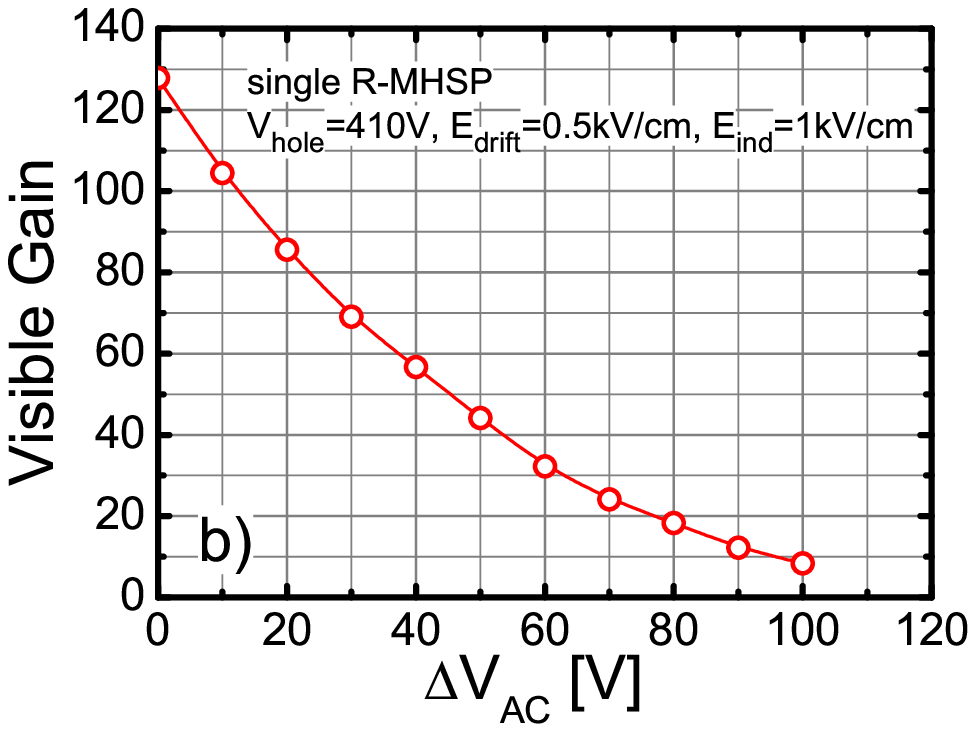}
} \subfigure {
    \label{fig:Fig.5:c}
    \includegraphics[width=7cm]{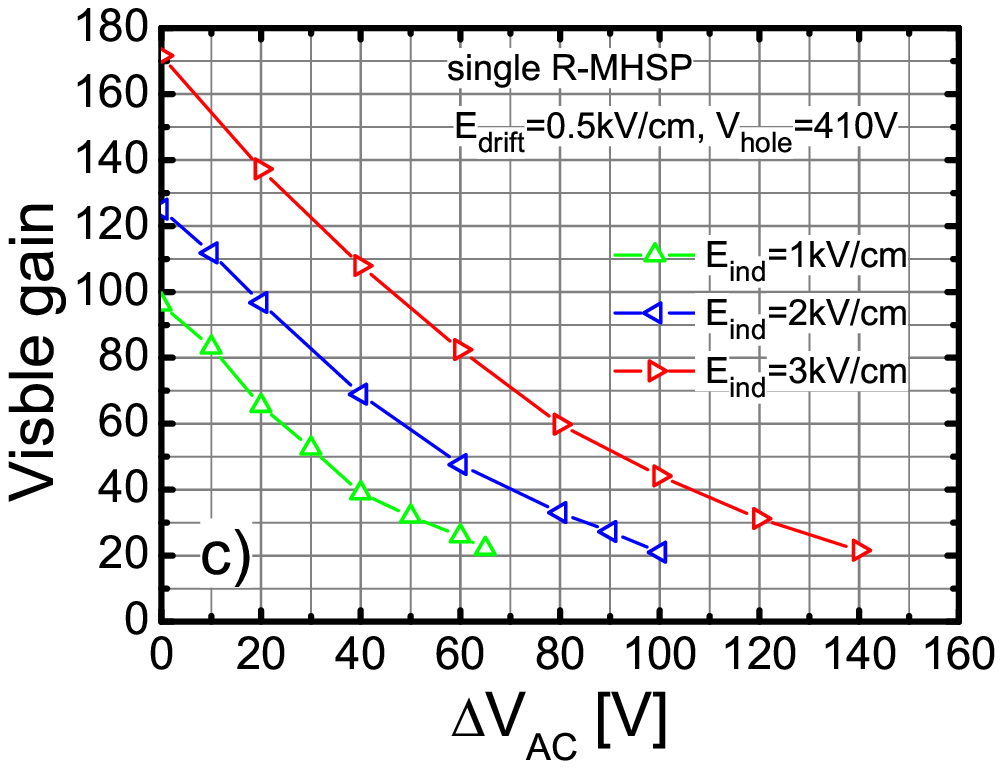}
} \caption{a) The electron currents measured in the R-MHSP setup of
\protect \ref{fig:Fig.3:a}, on the R-MHSP anode strips, $I_{RA}$,
and on the bottom electrode, $I_{BOT}$, as function of the
anode-to-cathode voltage difference, ${\Delta}V_{AC}$. The R-MHSP
avalanche gain within the holes is ${\sim}$500. The conditions are
given in the figure. b) The visible gain of a single R-MHSP as
function of anode-to-cathode voltage. c) Variation of the visible
gain of R-MHSP with ${\Delta}V_{AC}$ for different transfer field
values $E_{trans}$, measured with an R-MHSP of 50 micron holes.} 
\label{fig:Fig.5}
\end{center}
\end{figure}

Using MAXWELL 3D software package \cite{maxwell} we found for the
R-MHSP (\ref{fig:Fig.4}) that the field at the topside electrode, in
the immediate proximity of the holes entrance, is not influenced by
the voltage difference applied to the strips at the bottom-side
electrode. This means that the electron transport and focusing into
the R-MHSP holes is not affected by  ${\Delta}V_{AC}$ and should be
studied only as function of $E_{drift}$ and $V_{hole}$. As
emphasized above, for efficient single-photoelectron detection the
visible gain of the first R-MHSP should be carefully monitored.
Taking into account that when  ${\Delta}V_{AC}$ is raised, only a
few percents of the R-MHSP avalanche electrons are extracted to the
next element, and that single-electron multiplication is
exponential, we estimated that the first R-MHSP element should have
a visible gain $>$20 in order to detect photoelectrons with good
efficiency; the exact detection efficiency is a subject for future
study, with pulse-counting method. Our current estimation is based
on the avalanche-size distribution in single electron
multiplication, which in general should follow Polya formula
\cite{vavra93,vavra96,arley}, but in the case of hole
multiplication, in most gases \cite{buzu00}, simplifies to an
exponential distribution: $P(q)=exp(-q/\overline{q})$, $q$ and
$\overline{q}$ are the number and the mean of electrons in a
single-electron induced avalanche. With $\overline{q}$ being the
visible gain, $G_{VIS}$, of the R-MHSP, we can calculate that for
$G_{VIS}$=20-25 there is 90-92$\%$ chance to have two electrons or
more reaching the element following the R-MHSP, and these have high
probability to be focused into its holes and be further multiplied
and detected.

\ref{fig:Fig.5} shows the currents measured (setup of
\ref{fig:Fig.3:a}) on the R-MHSP anode strips $I_{RA}$, and on the
bottom electrode $I_{BOT}$, as function of the anode-to-cathode
strips voltage, ${\Delta}V_{AC}$. $V_{hole}$ was kept constant at
410V, resulting in avalanche gain in the holes $\sim$500, as derived
from $(I_{RA}+ I_{BOT})/I_{PC0}$ (the electron current on the
cathode strip was nearly zero). $E_{ind}$ and $E_{drift}$ were kept
at 1 and 0.5kV/cm, respectively. While the sum of both currents is
constant, varying ${\Delta}V_{AC}$ changes the current sharing
between these two electrodes, namely the charge extraction from the
R-MHSP holes to the next element (i.e. change of visible gain). As
much as it is desirable to increase ${\Delta}V_{AC}$ to divert more
ions to the cathode strips, the drop in visible gain
(\ref{fig:Fig.5:b}) sets a limitation on this variable, at about
70V, corresponding to visible gain $\sim$25. This value could be
further raised if the loss of electrons could be compensated by a
further increase of $V_{hole}$, but from our experience
$V_{hole}$=410V is already quite a high, though safe, operation
voltage. Another possibility to maintain visible gain $\sim$25 with
higher ${\Delta}V_{AC}$ is by raising $E_{ind}$, as depicted in
\ref{fig:Fig.5:c}. The data in \ref{fig:Fig.5:c} confirms the well
known \cite{bachmann99,dirk04b} linear dependence of the visible
gain on the induction field. It should be noted that the data of
\ref{fig:Fig.5:c} only were taken with an R-MHSP of smaller holes
(50 microns), which explains the minor difference in visible gain
compared to \ref{fig:Fig.5:b}.

\begin{figure}[!h]
\begin{center}
\renewcommand{\thesubfigure}{\thefigure\alph{subfigure}}
\makeatletter
\renewcommand{\p@subfigure}{Fig.\space}
\makeatother
\subfigure {
    \label{fig:Fig.6:a}
    \includegraphics[width=9cm]{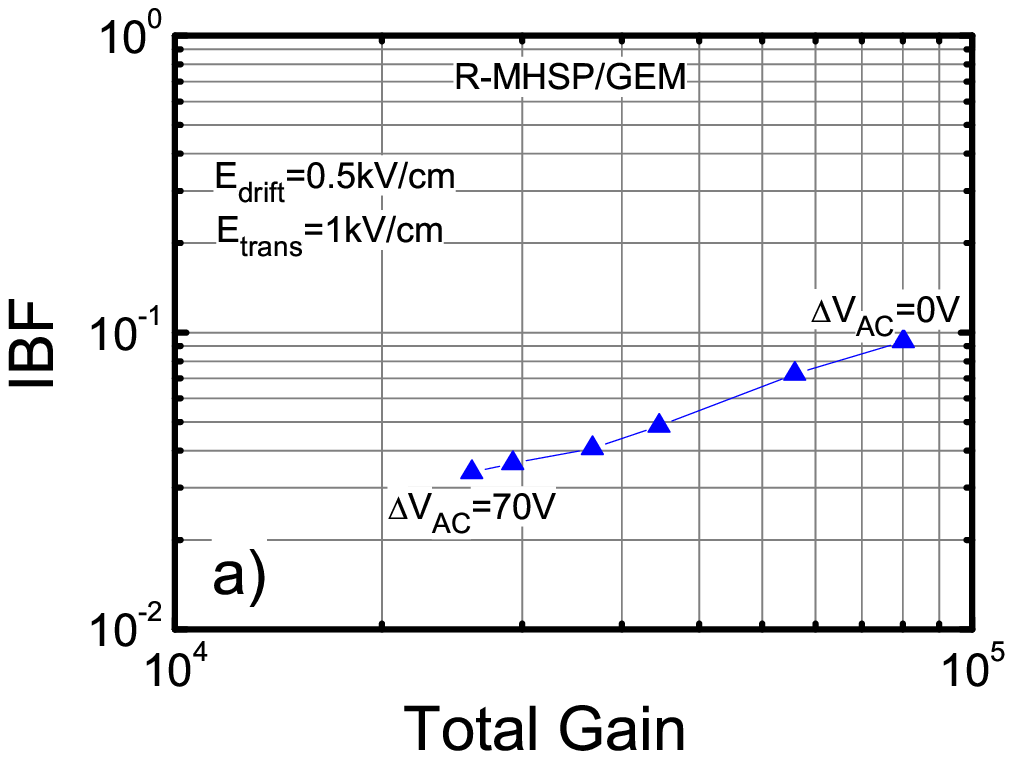}
}
\subfigure {
    \label{fig:Fig.6:b}
    \includegraphics[width=9cm]{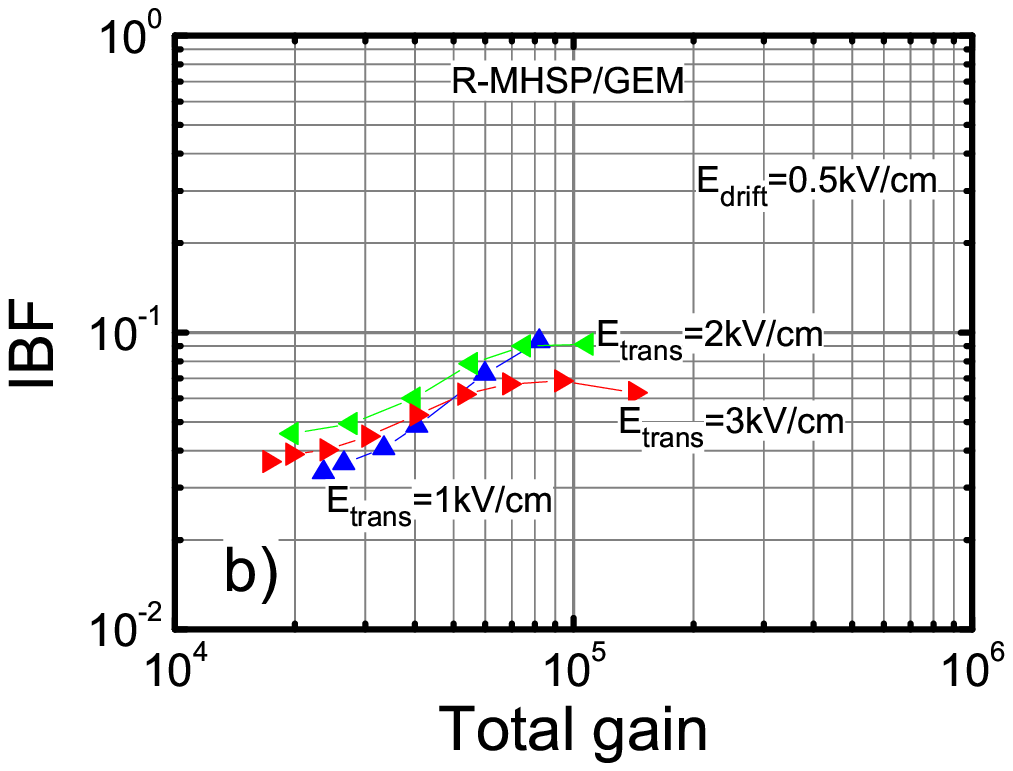}
}
\caption {a) The IBF measured in the R-MHSP/GEM structure of
\protect \ref{fig:Fig.3:b}, as a function of the absolute total
gain. The latter was varied by changing the voltage difference
across the R-MHSP strips, ${\Delta}V_{AC}$. The conditions are given
in the figure. b) Variation of IBF with transfer field. The total
gain was varied by changing the voltage difference across the R-MHSP
strips, ${\Delta}V_{AC}$: ${\Delta}V_{AC}$ was varied from 0V to 70,
100 and 140V, for $E_{trans}$=1, 2 and 3kV/cm, respectively. The
gain of the GEM was kept constant at ${\sim}$2000 in both cases.} 
\label{fig:Fig.6}
\end{center}
\end{figure}

The IBF reduction capability of the R-MHSP was studied
(\ref{fig:Fig.3:b}) with an R-MHSP followed by a GEM, of which the
avalanche acts as the source of back-flowing ions. The R-MHSP was
biased at $V_{hole}$= 410V; the GEM was biased at 420V (gain
$\sim$2000); the transfer field in the gap between them was
$E_{trans}$=1kV/cm and the drift field was $E_{drift}$=0.5kV/cm. The
total avalanche current in this configuration was measured as the
sum of currents from the bottom anode and from the bottom GEM
electrode. The IBF was calculated as the ratio of the PC current
under multiplication, to the total avalanche current.
\ref{fig:Fig.6:a} shows the IBF measured in these conditions, as
function of the total gain of both elements; the latter was adjusted
only by varying the voltage difference between the R-MHSP strips
(${\Delta}V_{AC}$). The IBF was reduced by a factor of 3 and the
visible gain was reduced by a factor of 5 while raising
${\Delta}V_{AC}$ from 0V (GEM-like mode) to 70V. In
\ref{fig:Fig.6:b}, the same IBF variation caused by raising
${\Delta}V_{AC}$, is shown for different transfer fields. With a
higher transfer field the visible gain is higher and
${\Delta}V_{AC}$ can be further raised: thus at $E_{trans}$=1kV/cm,
${\Delta}V_{AC}$ was varied from 0V to 70V while at $E_{trans}$=2
and 3kV/cm, ${\Delta}V_{AC}$ was raised to 100 and 240V,
respectively; the maximal ${\Delta}V_{AC}$ values correspond to
R-MHSP visible gain of $\sim$25. However, as is obvious from
\ref{fig:Fig.6:b}, the IBF for different transfer fields is rather
similar, and the best performance is with $E_{trans}$=1kV/cm.

\subsection{R-MHSP/2GEM/MHSP cascaded multiplier.}

        The R-MHSP/2GEM/MHSP cascaded-multiplier configuration
shown in \ref{fig:Fig.3:c} was chosen based on the reduced IBF with
an MHSP as a last-stage MHSP \cite{maia04} and its further potential
reduction by an R-MHSP as a first element. Note, however, that their
optimization are opposite and although in both operation modes the
IBF reduces with increased strips voltage difference, in the MHSP
this increase means higher total gain while in the R-MHSP it means
lower visible gain. The optimized fields configuration suggested in
\cite{maia04} was combined with the insight from the R-MHSP study
above, and we chose the following parameters (see
\ref{fig:Fig.3:c}): $E_{drift}$=0.5kV/cm; $E_{trans1}$=1kV/cm;
$E_{trans2}$ = $E_{trans3}$ =0.28kV/cm in order to suppress ion
transport (though at the expense of electron transport suppression)
and $E_{ind}$=-4kV/cm in order to collect ions at the bottom mesh
cathode. The first R-MHSP voltages were: $V_{hole1}$=410V;
${\Delta}V_{AC1}$=70V following the results above. The GEM voltages
were: $V_{GEM1}$=$V_{GEM2}$=280V, equivalent to a gain of ~60 on
each. The bottom MHSP voltages were: $V_{hole2}$=300V between the
top and cathode strip electrodes; we have varied ${\Delta}V_{AC2}$
on the bottom MHSP, to vary the total gain of the detector.

\begin{figure} [!h]
  \begin{center}
  \makeatletter
    \renewcommand{\p@figure}{Fig.\space}
  \makeatother
    \epsfig{file=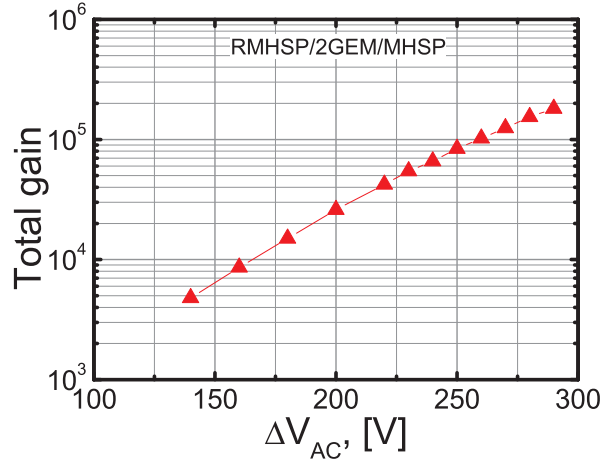, width=9cm}
    \caption{The total gain of the R-MHSP/2GEM/MHSP cascaded multiplier
of \protect \ref{fig:Fig.3:c}, as function of the anode-to-cathode
voltage difference, ${\Delta}V_{AC2}$, of the last MHSP element:
$E_{drift}$=0.5kV/cm, $V_{hole1}$=410V, ${\Delta}V_{AC1}$=70V,
$E_{trans1}$=$E_{trans2}$=0.28kV/cm, $V_{GEM1}$=$V_{GEM2}$=280V,
$V_{hole2}$=300V, $E_{ind}$=-4kV/cm.}
    \label{fig:Fig.7}
  \end{center}
\end{figure}

The total gain as function of ${\Delta}V_{AC2}$ on the bottom MHSP
is plotted in \ref{fig:Fig.7}. A charge multiplication of 50 was
achievable on the anode strips of the bottom MHSP, by raising the
strip voltage ${\Delta}V_{AC2}$ to 290V, in accordance with the
results of \cite{maia04}.

\begin{figure} [!h]
  \begin{center}
  \makeatletter
    \renewcommand{\p@figure}{Fig.\space}
  \makeatother
    \epsfig{file=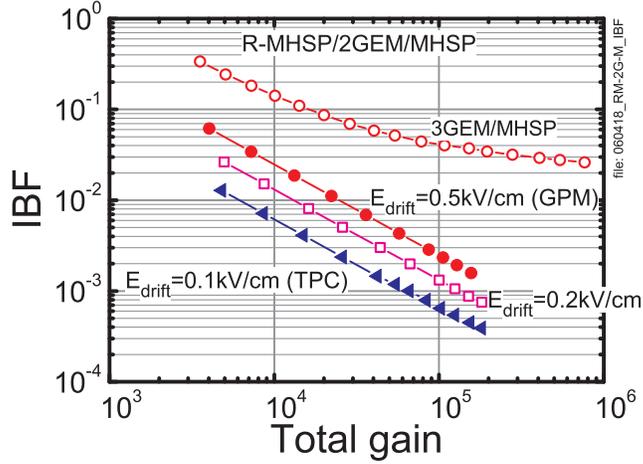, width=9.5cm}
    \caption{The IBF as a function of the total gain of the
R-MHSP/2GEM/MHSP cascaded detector, with a semitransparent
photocathode shown in \protect \ref{fig:Fig.3:c}; the IBF is plotted
for drift fields of 0.1-0.5 kV/cm. The operation conditions are the
same as in \protect \ref{fig:Fig.7}. Shown for comparison (open
circles) is the IBF for a 3GEM/MHSP detector with a reflective PC
\protect \cite{maia04}.}
    \label{fig:Fig.8}
  \end{center}
\end{figure}

The recorded IBF is presented in \ref{fig:Fig.8} as function of the
total gain, for drift-field values of 0.1, 0.2kV/cm (TPC conditions)
and 0.5kV/cm (GPM conditions). In the GPM-like operation mode, the
lowest IBF was 1.5*10$^{-3}$ at a detector total gain of 1.5*10$^5$.
It means that per a single-photon event, on the average 200 ions
reach the PC. The IBF recorded at the same total gain in the TPC
operation mode (drift field of 0.1kV/cm) is $\sim$4*10$^{-4}$, which
is $\sim$5 times lower, as expected from the linear dependence of
IBF on the drift field discussed in \cite{bondar03}. At a drift
field 0.2kV/cm the IBF is intermediate, of 7*10$^{-4}$ at a total
gain 1.5*10$^5$. Note that the IBF curves of the R-MHSP/2GEM/MHSP
cascaded-multiplier (\ref{fig:Fig.8}) drop as IBF=$C/G_{tot}$, $C$
being a constant, over the entire total gain range; this is in
contrast with the case of the 3GEM/MHSP multiplier with reflective
PC \cite{maia04}, shown for comparison in \ref{fig:Fig.8}, where the
IBF decreases slower than 1/$G_{tot}$ at total gains above 2*10$^4$.
Based on the definitions of IBF (IBF=$I_{PC}/I_{A}$) and total gain
($G_{tot}=I_{A}/I_{PC0}$), it is clear that $C=I_{PC}/I_{PC0}$,
which means that at a given drift field, the PC ion current is
constant and does not increase with ${\Delta}V_{AC2}$. In other
words, all ions created at the anode strips of the MHSP do not reach
the PC, which demonstrates full ion suppression by the electrode
cascade preceding the MHSP strips. On the contrary, with the
3GEM/MHSP detector an increasing fraction of ions from the anode
strips reach the PC, pointing at a poorer ion trapping capability of
this cascade.

\subsection{2R-MHSP/MHSP cascaded multiplier with a reflective PC.}

An important configuration of cascaded gaseous photomultipliers is
that with a reflective photocathode deposited on top of the first
multiplying element, a GEM \cite{dirk04b} or a THGEM
\cite{chechik05}. In addition to the higher QE values of reflective
photocathodes, in this configuration the sensitivity to
ionizing-particle background is considerably reduced \cite{dirk04b},
because the gaseous drift gap is practically eliminated. The best
IBF recorded so far in a reflective-PC GPM, of IBF=3$\%$, was in a
3GEM/MHSP configuration with the photocathode deposited on the top
GEM \cite{maia04}. We have investigated a few other solutions, among
them a double R-MHSP followed by a MHSP, with a CsI photocathode
deposited on the top R-MHSP \cite{breskin05} (\ref{fig:Fig.3:d}).

The drift gap above the photocathode, defined by a mesh electrode
placed at a distance of 3.5mm above the top surface of the R-MHSP
(\ref{fig:Fig.3:d}), was biased to have $E_{drift}$=0kV/cm, which
according to \cite{dirk04b,chechik05} provides the best
photoelectron collection efficiency from the photocathode into the
nearest hole. It was demonstrated that a slightly reversed field in
this gap repels negative charges deposited there and significantly
reduces the sensitivity to charged-particles background
\cite{phenix05}, at the expense of a minor (few percent) drop in
photoelectron detection efficiency \cite{dirk04b}. We studied the
IBF under these conditions, with $E_{drift}$=-0.1kV/cm. In our
experiment the potential across the hole of the first R-MHSP,
$V_{hole1}$=400V, ensured photoelectron extraction from the PC above
90$\%$; the anode-to-cathode strip voltage was set to be
${\Delta}V_{AC1}$=50V and the first and the second transfer fields
were set to rather high values, to provide a visible gain $>$20 for
the first R-MHSP element and further good electron transfer to the
last MHSP element:$E_{trans1}$=3kV/cm and $E_{trans2}$=5kV/cm; the
induction field was $E_{ind}$=-8kV/cm; the second R-MHSP and the
last MHSP were biased at rather high values of $V_{hole2}$=450V and
$V_{hole3}$=430V across the holes, respectively. This however,
allowed setting ${\Delta}V_{AC2}$ at the second R-MHSP to 100V for
better trapping of ions from the last MHSP.

\begin{figure} [!h]
  \begin{center}
  \makeatletter
    \renewcommand{\p@figure}{Fig.\space}
  \makeatother
    \epsfig{file=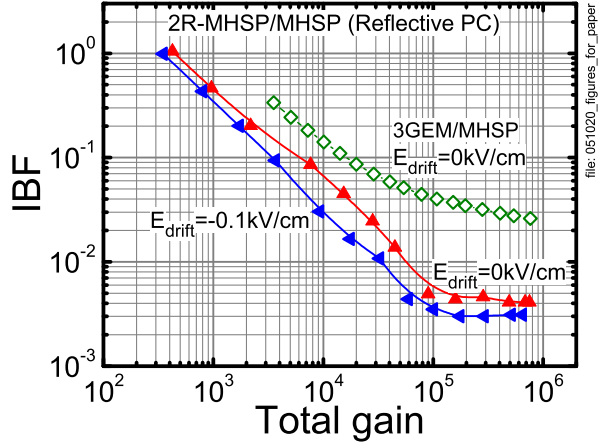, width=9cm}
    \caption{The IBF as a function of the total gain for the
2R-MHSP/MHSP detector (of \protect \ref{fig:Fig.3:d}) with
reflective photocathode on first R-MHSP top electrode, for two
different drift fields. Experimental conditions: $V_{hole1}$=400V,
${\Delta}V_{AC1}$=50V, $E_{trans1}$=3kV/cm, $V_{hole2}$=450V,
${\Delta}V_{AC2}$=100V, $E_{trans2}$=5kV/cm, $V_{hole3}$=430V,
$E_{ind}$=-8kV/cm. Shown for comparison (open diamonds) is the IBF
for a 3GEM/MHSP detector with a reflective PC \protect
\cite{maia04}.}
    \label{fig:Fig.9}
  \end{center}
\end{figure}

The IBF in the 2R-MHSP/MHSP detector with reflective PC is shown on
\ref{fig:Fig.9} for two different drift fields: 0 and -0.1kV/cm
(reversed field). The lowest IBF recorded is 3*10$^{-3}$ in the
total gain range of 10$^5$ - 7*10$^5$ (\ref{fig:Fig.9}). The plots
for both drift fields are nearly the same and start saturating at
total gain $\sim$10$^5$. This, to our opinion, is due to non optimal
operation conditions: 1) a too low ${\Delta}V_{AC1}$ at the first
R-MHSP and 2) a too high $E_{trans2}$ compared to the value of
0.2-0.5kV/cm in \cite{maia04}. However, the IBF values achieved in
the present study are by an order of magnitude better than the best
value obtained so far with 3GEM/MHSP detector + reflective PC
\cite{maia04}, at a total gain of 10$^5$, shown for comparison in
\ref{fig:Fig.9}.

\section{Conclusions and discussion.}

In this work we have continued our long ongoing studies of IBF
reduction in cascaded electron multipliers, searching for further
improvements that will permit their operation with a visible-light
sensitive (e.g. bialkali) PC. Following the 5-fold IBF reduction of
the MHSP as a last cascade-element \cite{maia04} and the preliminary
results with the reversed-bias R-MHSP as a first cascade-element
\cite{veloso05}, we further studied the ion trapping properties of
the R-MHSP. These were systematically investigated with R-MHSP as a
stand-alone element, as a first element in a R-MHSP/2GEM/MHSP
cascaded multiplier and in a reflective-PC gaseous photomultiplier
based on 2R-MHSP/MHSP multiplier.

We found that raising the anode-to-cathode voltage in the R-MHSP
from 0V (GEM mode) to -70V (R-MHSP mode) led to a 3-fold reduction
of the IBF. We further evaluated the associated reduction in the
visible gain of the R-MHSP first element and we argue that it can be
maintained sufficiently high to ensure good detection efficiency of
single-electron events: with hole multiplication $\sim$500, a
transfer field 1kV/cm and anode-to-cathode voltage -70V the visible
gain is 25, implying that in 92$\%$ of the events at least two
electrons are transferred to the following element in the cascade.

The R-MHSP was investigated in an R-MHSP/2GEM/MHSP and 2R-MHSP/MHSP
detector configurations, combining a CsI photocathode operated in
semitransparent and reflective mode correspondingly. For the
R-MHSP/2GEM/MHSP with a semitransparent PC, the drift field between
the photocathode and the R-MHSP was selected to simulate GPM and TPC
operation conditions. For the 2R-MHSP/MHSP detector with reflective
PC, the drift field between the PC and the mesh was set either at
0kV/cm (corresponding to maximum photo-electron extraction from the
PC) or slightly reversed (-0.1kV/cm), in order to decrease the
detector's sensitivity to charged-particles background.

\subsection{GPM with semitransparent PC:}

In the semitransparent-PC R-MHSP/2GEM/MHSP GPM, the IBF was found to
be inversely proportional to the total gain; its best value in
GPM-like mode ($E_{drift}$=0.5kV/cm), compatible with good
single-electron detection efficiency, was 0.15$\%$ at a total gain
of 1.5*10$^5$. This IBF is too high for a stable operation of the
multiplier in combination with a visible-light sensitive
photocathode (e.g. bialkali), due to the high emissivity of these
photocathodes; for example, with a bialkali PC, the 225 ions
(1.5*10$^5$ x 0.015) impinging on it per single-photoelectron event
will induce on the average 8 secondary electrons \cite{dirk_tes,
dirk06}. Thus at least an 8-fold IBF reduction is further required
to bring the gain-limiting secondary avalanches to an acceptable
level.

A further IBF reduction will also prolong the photocathode's
lifetime, affected by the ion impact. At present the photocathode
lifetime of  the R-MHSP/2GEM/MHSP GPM can be estimated, for example,
from the recently measured aging rate of a bialkali PC: ~20$\%$ loss
of QE at an accumulated ion charge of 2$\mu$C/mm$^2$ on a K-Sb-Cs PC
\cite{breskin05,lyas06}. With a photon flux of 1kHz/mm$^2$ and a
QE=30$\%$, such charge will be accumulated during 7 years of
operation at a total gain of 10$^5$ and with IBF=2*10$^{-3}$.

It should be noted that the IBF reduction with the present
multiplier configuration is superior to that presented in our
previous work \cite{veloso05}; it is a result of a better-quality
MHSP electrodes and a combination of an R-MHSP and MHSP as first and
last elements, respectively.

\subsection{GPM with reflective PC:}

In the reflective-PC 2R-MHSP/MHSP GPM the IBF was found to be 7-5
times lower than in the previously reported reflective-PC 3GEM/MHSP
GPM \cite{maia04}, in the total gain range 10$^5$ - 10$^6$. It is in
general more difficult to block the ions when using a reflective PC
directly deposited on top of the hole-multiplier, because the field
values and directions on this surface are dictated solely by the
hole dipole filed, and under the large multiplication in the holes
(dictated by gain considerations) this field is rather high
\cite{dirk04b}. Nevertheless, the R-MHSP as a first (and second)
element indeed provides further IBF reduction.

\subsection{TPC operation mode:}

Due to the almost linear dependence of IBF on drift field, it is
5-10 times smaller in TPC operation mode compared to GPM one. In the
R-MHSP/2GEM/MHSP detector and with $E_{drift}$=0.1kV/cm, the IBF is
4*10$^{-3}$ and 4*10$^{-4}$ at total gas gains of 1.5*10$^4$ and
1.5*10$^5$, respectively. It means that per single-electron
multiplication event, $\sim$60 ions flow back to the drift region,
independent of the gain.

The present study was motivated by the need to reduce the IBF in
GPMs and therefore the structures under study were primarily
adjusted for single-photoelectron detection conditions at high total
gain ($\sim$10$^5$). However, the majority of IBF studies
\cite{bondar03,lotze06,sauli05} with multi-GEM detectors were
carried out at lower gains (10$^4$), oriented towards TPC operation.
With the presently studied multiplier configurations and fields, the
IBF value in TPC-mode and at total gain $\sim$10$^4$ is practically
not improved compared to the best results obtained so far with
asymmetric 3-GEM cascade \cite{lotze06}. However, we expect that a
better optimization of the R-MHSP/2GEM/MHSP multiplier in TPC-mode
conditions (total gain 10$^4$ and $E_{drift}$=0.1kV/cm) will further
reduce the IBF. On the contrary, at total gain 10$^5$, one order of
magnitude improvement was achieved, compared to \cite{maia04} as
discussed above, and compared to \cite{bondar03} with a 3-GEM
detector and $E_{drift}$=0.5kV/cm, reaching IBF=0.02 at total gain
6*10$^4$.

An idea to further reduce the IBF might be the use of single, or
several R-MHSPs with anode and cathode strips patterned on both
faces. According to our preliminary studies of an R-MHSP mounted
with the strips facing the drift region (flipped R-MHSP), an
increase of the anode-to-cathode voltage difference did not affect
much the R-MHSP gain in the holes; in addition, the cathode strips,
now facing the drift region, seem to efficiently capture the
back-flowing ions from successive cascade elements. The results of
these studies will be reported in another article.

\acknowledgments

This work is partly supported by the Israel Science Foundation,
grant No 402/05, by the MINERVA Foundation and by Project
POCTI/FP/63441/2005 through FEDER and FCT (Lisbon). A. Breskin is
the W.P. Reuther Professor of Research in The Peaceful Use of Atomic
Energy.

\end{document}